\documentclass[runningheads]{llncs}
\usepackage{graphicx}
\usepackage{xcolor}
\usepackage{float}
\usepackage{amsmath}
\usepackage{soul}

\newcommand{\repeatthanks}{\textsuperscript{\thefootnote}}

\begin{document}
\title{Neural parameters estimation for brain tumor growth modeling}
\author{Ivan Ezhov\thanks{The authors contributed equally to the work.}\inst{1} \and Jana Lipkova\repeatthanks\inst{1} \and Suprosanna Shit\inst{1} \and Florian Kofler\inst{1} \and Nore Collomb\inst{2} \and Benjamin Lemasson\inst{2} \and Emmanuel Barbier\inst{2} \and Bjoern Menze\inst{1}}
\institute{Department of Informatics, Technical University of Munich \and Grenoble Institut des Neurosciences, University Grenoble Alpes \\ \email{ivan.ezhov@tum.de}}
\authorrunning{Ivan Ezhov et al.}
\maketitle              
\begin{abstract}
Understanding the dynamics of brain tumor progression is essential for optimal treatment planning. Cast in a mathematical formulation, it is typically viewed as evaluation of a system of partial differential equations, wherein the physiological processes that govern the growth of the tumor are considered. To personalize the model, i.e. find a relevant set of parameters, with respect to the tumor dynamics of a particular patient, the model is informed from empirical data, e.g., medical images obtained from diagnostic modalities, such as magnetic-resonance imaging. Existing model-observation coupling schemes require a large number of forward integrations of the biophysical model and rely on simplifying assumption on the functional form, linking output of the model with the image information. In this work, we propose a learning-based technique for the estimation of tumor growth model parameters from medical scans. The technique allows for explicit evaluation of the posterior distribution of the parameters by sequentially training a mixture-density network, relaxing the constraint on the functional form and reducing the number of samples necessary to propagate through the forward model for the estimation. We test the method on synthetic and real scans of rats injected with brain tumors to calibrate the model and to predict tumor progression.
\end{abstract}
\section{Introduction}
Modeling brain tumor progression holds a promise of optimizing clinical treatment planning. An appropriate tumor model, personalised with respect to the patient-specific growth dynamics, could quantify clinically relevant information - the tumor's morphology and its character of evolution \cite{ref_article11,ref_article3}. Existing mathematical description of the pathophysiological system spans from the intracellular level of gene expression to the macroscopic level of bio-mechanical tumor-tissue interaction. The latter is the scale at which the medical imaging analysis is typically carried out as this is the scale where medical scans are most interpretative. 

Among the family of macroscopic models, the reaction-diffusion class of equations \cite{ref_article2} is most widely adopted to characterize information visible on medical scans. Under such equations the evolution of tumor cell density is tracked by considering tumor-relevant physiological processes, such as proliferation of cancerous cells, i.e. increase of the cells number due to its division, and the cells' migration into surrounding tissue. Various approaches have been developed to link the output of the model, the distribution of the cell density, with the tumor visible on images \cite{ref_article11,ref_article3,ref_article4,ref_article5,ref_article7,ref_article8,ref_article9,ref_article10}. 
Methods as in \cite{ref_article5} make a certain assumption on the cell density along visible tumor outlines and fit the model output to image observation that includes lesion growth and tissue displacement. The model adjustment, realized by means of a PDE-constrained optimisation scheme, allows to obtain a point estimate of free model parameters. Bayesian methods \cite{ref_article11,ref_article8,ref_article9,ref_article10} cast the problem in a probabilistic formulation and provide estimation of the parameters along with confidence intervals via Markov Chain Monte Carlo (MCMC) sampling. The authors of \cite{ref_article8} rely on the travelling wave formulation of \cite{ref_article4} together with a Bayesian parameter estimation. In \cite{ref_article11,ref_article10}, authors construct a probabilistic graphical model wherein the probability of imaging signal is defined to be dependent on the biophysical model's output. For magnetic-resonance images (MRI), the probability of observing abnormality is defined as a logistic sigmoid function of the tumor cell density. Phenomenological introduction of the functional form leaves the question whether it possesses a capacity to approximate the mapping between cell density and the imaging information. Also, generating samples from the posterior distribution as with the MCMC methods requires large number of evaluations of the forward model, which can be of the order 10-100 thousand evaluations \cite{ref_article11,ref_article10}. This results in an expensive computational cost, impeding clinical validation of more complex models and eventually the approach's adoptability to a routine daily use within clinical settings. 

In this paper, we adopt methodological advances in the estimation of forward model parameters, relying on learning-based strategy \cite{ref_article12,ref_article13}. The technique allows for explicit evaluation of the distribution over the parameters by training a mixture-density network (MDN) \cite{ref_article14}. The MDN, modeled as a feedforward fully-connected network, maps the output of the model to parameters of the distribution in a non-linear fashion. As theoretical works \cite{ref_article15} prove, such a network can serve as a universal function approximation, thus relaxing the necessity of introducing an explicit form for the likelihood, relating the model output and imaging data. In summary, the contributions of this paper are threefold: (1) We make the technique applicable to PDE-based tumor growth models, (2) We validate our method on synthetic and real data of rats implanted with cancer cell lines, using two time points for the model initialization and calibration, (3) We demonstrate that the technique provides more accurate parametric estimations and requires less forward model's samples as compared to explicit Bayesian formulation even with highly efficient MCMC sampling method.   
\section{Method}
\subsubsection{Tumor growth model.}
We base our forward model on the reaction-diffusion equation, describing the tumor progression via spatial and temporal evolution of the cancerous cell density. Particularly, a special type of the reaction-diffusion formalism, the Fisher-Kolmogorov equation, is used:  
\begin{equation}
\frac{\partial u}{\partial t}= \nabla (\textbf{D}\nabla u)+\rho u(1-u), \quad in \quad\Omega \\
\end{equation}
\begin{equation}
\textbf{n} \cdot \nabla u = 0, \quad in \quad\Gamma_{\Omega}.
\end{equation}
Eq. (1) considers two pathophysiological processes: the logistic proliferation of the cells and its diffusion into neighbouring tissue. $u$ denotes the tumor cell density in the volume of the brain $\Omega$, \textbf{D} is the diffusion tensor and $\rho$ denotes tumor proliferation rate. The diffusion is assumed to be heterogeneous: with different degree of infiltration in the white and the grey matters, and restricted in the ventricles area. We impose no-flux boundary condition Eq. (2), \textbf{n} denotes the unit vector orthogonal to the boundary of the simulation domain $\Gamma_{\Omega}$ and $\nabla$ is the gradient operator. We performed experiments with two variants of the model initialization: as a seed point at a fixed location \pmb{r}* ($u(\pmb{r},0)=u_0$ if $\pmb{r}=\pmb{r*}$, $u(\pmb{r},0)=0$ elsewhere), and as an approximation of the cell density distribution, obtained from an image observation at the first monitoring time point ($u(\pmb{r},0)=u_0(\pmb{r})$).
\subsubsection{Linking tumor model and image observation.}We calibrate the model parameters from image observations in the form of 3D binary tumor segmentations, obtained from MRI modalities. Two MRI modalities, T1-gadolinium hyper-intensities (featuring active tumor core) and T2-hyper-intensities (featuring whole tumor), were used in order to better describe the right tumor morphology. To make the output of the model consistent with the segmentations we make a physiologically plausible assumption that regions of abnormalities visible on the images correspond to regions of high cell infiltration. Respecting the assumption, we introduce two additional parameters $u^{T1},u^{T2}$ for thresholding the simulated cell density profile, leading to isolines of the tumor cell density that we assume to match outlines of the tumor visible in a given modality. The thresholded binary volumes are combined by element-wise summation to form a 3D label map.
\subsubsection{Neural parameters inference.}
We can view the forward model's output $X$ -- the 3D label map -- as a sample from a likelihood distribution $p(X | \theta)$ conditioned on a set of parameters $\theta=\{D,\rho,u^{T1},u^{T2}\}$. The distribution $p(X | \theta)$ cannot be in general evaluated, but its samples are readily available from the tumor model. Given an observation $X_{obs}$ -- segmentations of the tumor in the MRI modalities (summed element-wise), our goal is to infer the posterior distribution of the tumor model parameters, using the Bayes rule: $p(\theta | X_{obs})$ $\propto p(X_{obs} | \theta)p(\theta)$.

\begin{figure}[H]
\includegraphics[width=1.0\textwidth]{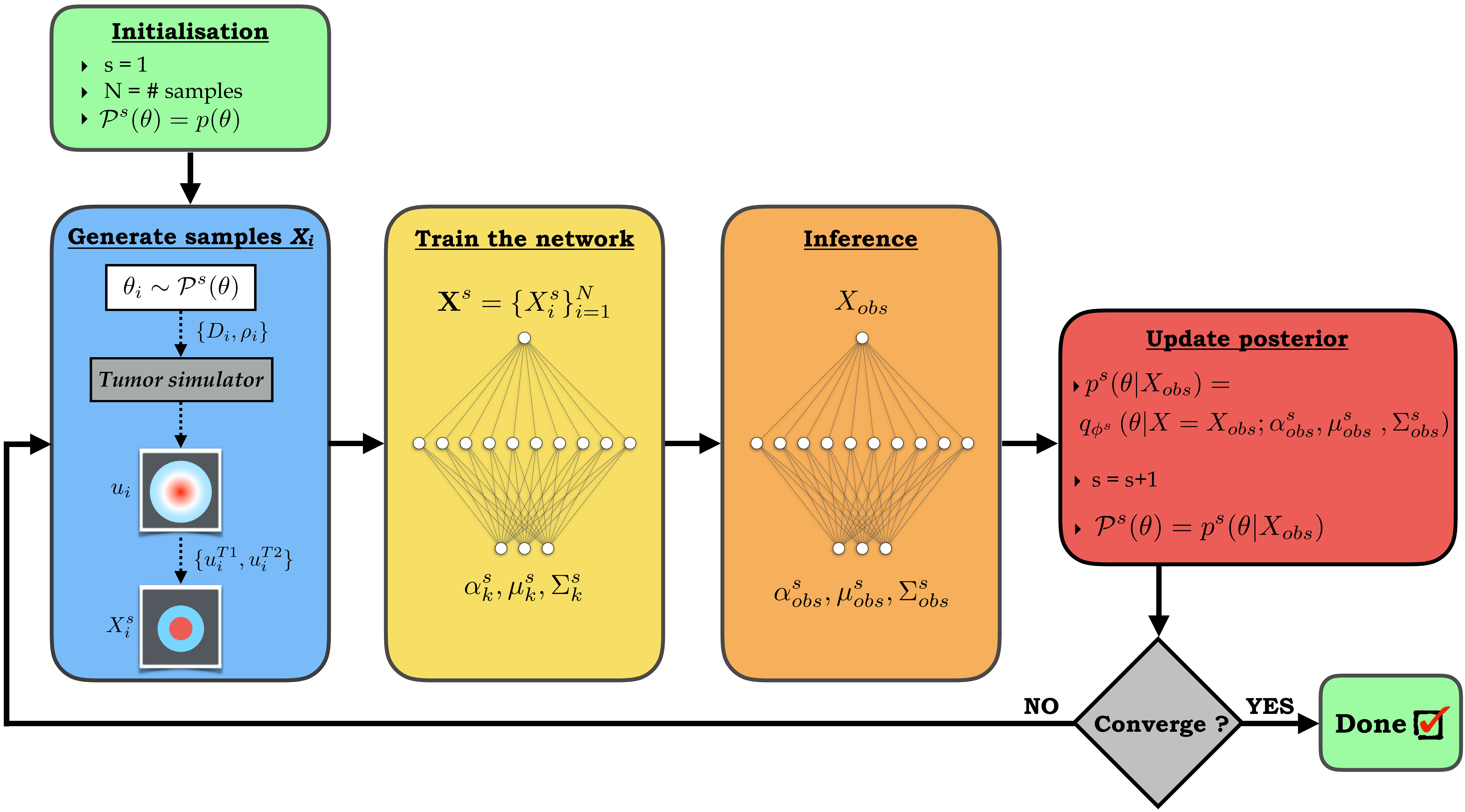}
\caption{\small{The neural posterior estimator. At the heart of it is a mixture density network that maps input data to closed form estimates of the model parameters. It guides the patient-specific simulation of the tumor growth in an efficient iterative fashion.}} \label{fig1}
\end{figure}

In \cite{ref_article11,ref_article10}, the likelihood is approximated by Bernoulli distribution with the parameter of the distribution defined as logistic sigmoid function. In our work, for inference of the forward model's parameters, we adopt a methodology that allows to learn a nonlinear mapping from the output of the model directly to posterior distribution over its parameters \cite{ref_article12}. The inference is based on the neural posterior estimation (NPE), wherein an approximated posterior $q_\phi(\theta|X)$, modeled as a mixture density network, converges to the true posterior $p(\theta | X)$ (via the Kullback-Leibler divergence minimization) by iteratively performing the following steps illustrated in Fig. \ref{fig1}: 

1) [\textit{Blue box}] Pairs $\{\theta_i,X_i \}_{i=1}^N$ are generated to form a training dataset. First, the \ul{tumor simulator} parameters $\{D_i,\rho_i\}$ are sampled from a prior $p(\theta)$ distribution (which is uniform at the first iteration step $s=1$) and corresponding simulation is propagated until the fixed time point $t^*$ to obtain the 3D cell density profile $u_i$. Then, $u_i$ is transformed to obtain binary segmentation masks, using the other two sampled parameters $\{u_i^{T1},u_i^{T2}\}$. Together, the segmentation masks form $X_i$.

2) [\textit{Yellow box}] The MDN is trained by taking $X_i$ as input and outputting parameters $\alpha^s_k, \pmb{\mu}^s_k, \pmb{\Sigma}^s_k$ of a \ul{mixture of Gaussians} $q_{\phi^s}(\theta|X)=\sum_k \alpha^s_k N(\theta|\pmb{\mu}^s_k,\pmb{\Sigma}^s_k)$ of K components. The objective of the approximated posterior $q_{\phi^s}(\theta|X)$ training is to maximize the total log-loss, $L({\phi^s}) = \sum_i log(q_{\phi^s}(\theta_i|X_i))$.

3) [\textit{Orange box}] The trained MDN is used to \ul{infer observation specific parameters} of the Gaussian mixture $\alpha^s_{obs}, \pmb{\mu}^s_{obs}, \pmb{\Sigma}^s_{obs}$ by evaluating $q_{\phi^s}(\theta |X=X_{obs})$ 
at the observation $X_{obs}$ - the label map, obtained from MRI segmentations at $t^*$.

4) [\textit{Red box}] Finally, the observation specific parameters are used to \ul{update estimation of the posterior} $p^s(\theta | X_{obs}) = q_{\phi^s}(\theta |X=X_{obs}; \alpha^s_{obs}, \pmb{\mu}^s_{obs}, \pmb{\Sigma}^s_{obs})$, which then used as a proposal distribution for sampling model parameters during the next iteration.
During successive iterations all the four steps are identical except the step 2, that requires modification of the training objective. To compensate for the fact that we sample from the proposal distribution, the objective function is weighted by a ratio between the prior and proposal distributions $ p(\theta_i)/p^s(\theta_i | X_{obs})$ \cite{ref_article12}:
\begin{equation}
L({\phi^s}) = \sum_i \frac{p(\theta_i)}{p^s(\theta_i | X_{obs})}log(q_{\phi^s}(\theta_i|X_i)) 
\end{equation}
\subsubsection{Implementation.}
We implement the tumor simulator using 3D extension of the multi-resolution adaptive grid solver \cite{ref_article16}, allowing for high-parallelization. The typical execution time is 20-40 seconds with 8 CPU cores. The architecture of the neural estimator represents a feedforward fully-connected network with a single hidden layer of 100 units with $tanh$ as an activation function. We initialize the weights of the network with He-normal \cite{ref_article18} at the first iteration and use the weights trained at the iteration step s for initialization at the s+1 step. The latter allows to implicitly reuse the samples from previous iterations: for memory efficiency, it is desirable to avoid having to store and reuse directly old samples since we are dealing with 3D volumes. The network was trained using the Adam optimizer \cite{ref_article17} for 100 epochs at each iteration. For each subject \textit{a separate network is employed}. We run the experiments on NVIDIA Quadro P6000 GPU.
\section{Experiments}
\subsubsection{Data.}
In our experiments, we use synthetic and real data of human glioma cells injected in the rat brain. For producing the synthetic data, we simulate a 3D tumor in the anatomy, obtained from rats brain atlas. We initialize the tumor as a seed point at a fixed location with the diffusion coefficient in the white matter $D_w=0.02\ [mm^2/day]$ greater than in the grey matter $D_w=10D_g$, and proliferation rate $\rho=0.6\ [1/day]$. To generate tumor segmentations masks, we threshold the simulated normalized cell density profile at $u^{T1}=0.7$ and $u^{T2}=0.25$ for T1 and T2 modalities, respectively, at a single calibration time point ($t^*$ = day 9). The day 11 is used for the model validation.
The real data were obtained by injecting F98 tumor cell lines in rats brain. The tumor progression was monitored at several time points from day 9 to day 16, using T1w, T2w, and DWI imaging modalities. The images were expert-annotated. Since the initial condition of tumor location and shape is unknown in the real rats due to the injection, we made use of the DWI modality at the first monitoring time point (day 9) for model initialization. The apparent diffusion coefficient (ADC), calculated from the DWI, can be considered to be inversely proportional to the tumor cell density \cite{ref_article19}. We used the ADC, confined within the T2w segmentation volume, as initial condition (in the late time states of tumor progression, the complex tumor microenviroment, hypoxia and necrosis complicate the simple inversely proportional relation). The binary segmentations from the T1w and T2w at the next time point ($t^*$ = day 11) are used for inference of the model parameters, and at the following days (14 and 16) we validate the model predictions, Fig. \ref{fig3}. 
\subsubsection{Results on synthetic data.}
\begin{figure}[H]
\centering\includegraphics[width=1.0\textwidth]{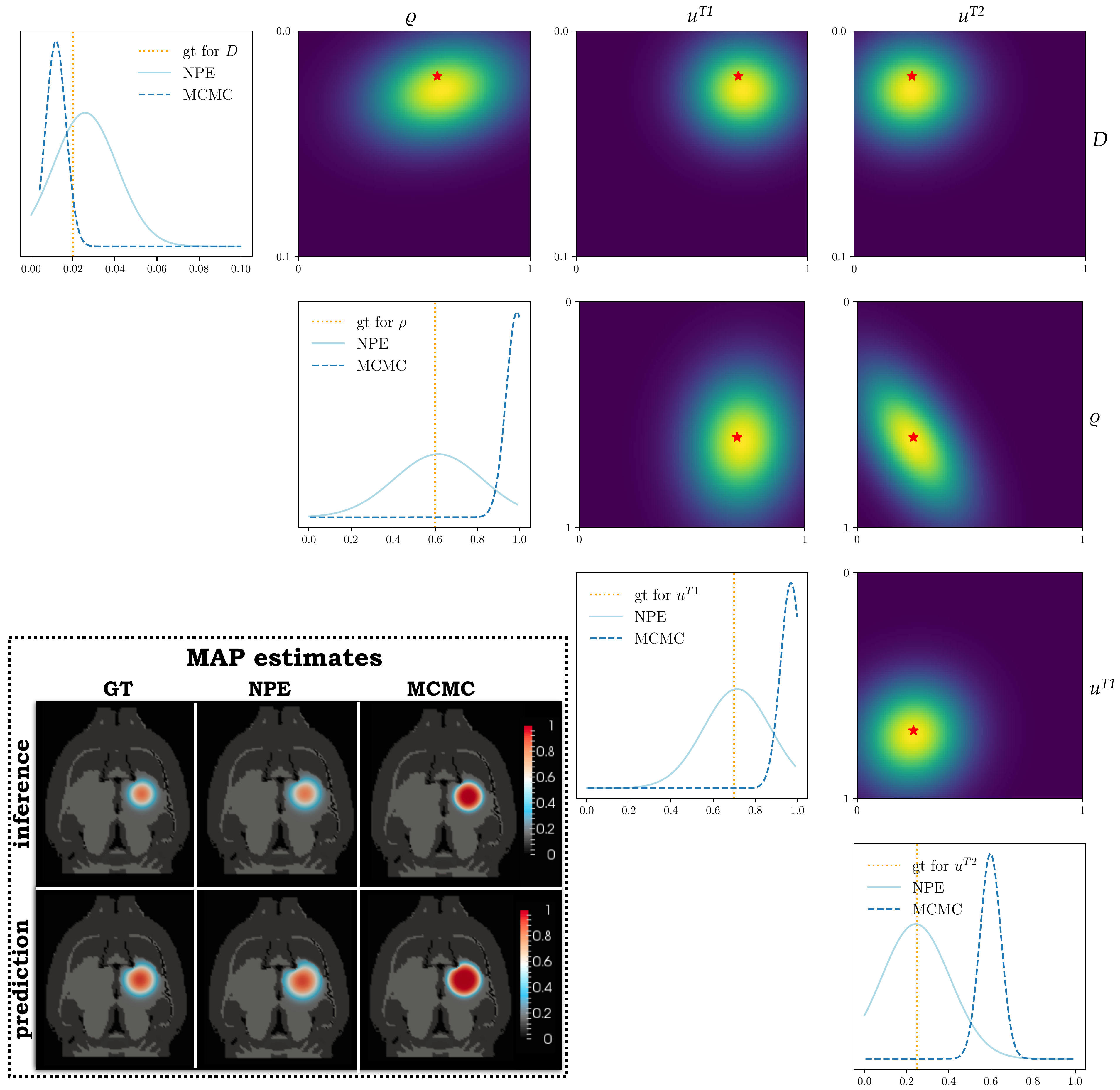}
\caption{\small{Posterior distribution of the tumor growth model's parameters inferred for the synthetic rats data: 1D distributions along the diagonal (for the NPE and MCMC methods) and 2D marginals (for the NPE) elsewhere. Depicted by red stars and orange vertical lines are ground truth data. Tumors simulated in the rats brain atlas using the ground truth parameters, and MAP parametric estimates obtained by the NPE and MCMC-based methods are shown on the inset. The top row depicts 2D slices of the cell density profile at the inference time point (day 9), and the bottom row - at the prediction time point (day 11).}} \label{fig4}
\end{figure}

For a sensitivity analysis of the inference, experiments on the synthetic data were first performed. In Fig. \ref{fig4} we show a pairwise correlation of the forward model parameters $\{D,\rho,u^{T1},u^{T2}\}$ obtained with the neural posterior estimator and the explicit Bayesian inference with MCMC sampling from \cite{ref_article11}. For both methods, 1000 samples were used for the inference. Depicted by red stars and orange vertical lines are ground truth (gt) data. The proposed method provides the maximum a posteriori estimation (MAP) 
\noindent for all the parameters in a close agreement with the gt data. In consistency with \cite{ref_article11}, for the MCMC, based on the likelihood formulation as a logistic sigmoid, the information in the form of binary segmentations is not sufficient to recover the gt parameters. At the same time, the NPE is more computationally efficient, since we observe accurate estimates after running 4 iterations of the posterior update, whereas the MCMC-based method requires about 20 sampling generations for convergence. This is attributed to a) sampling from a range of the parametric space more relevant to the observation after each iteration, b) efficient use of the samples as the technique does not imply any rejection thereof. The inset on the Fig. \ref{fig4} shows the tumor cell density computed with the MAP parametric estimations from each method, in comparison with the ground truth data.
\subsubsection{Results on real data.} 
\begin{figure}[H]
\centering\includegraphics[width=1.0\textwidth]{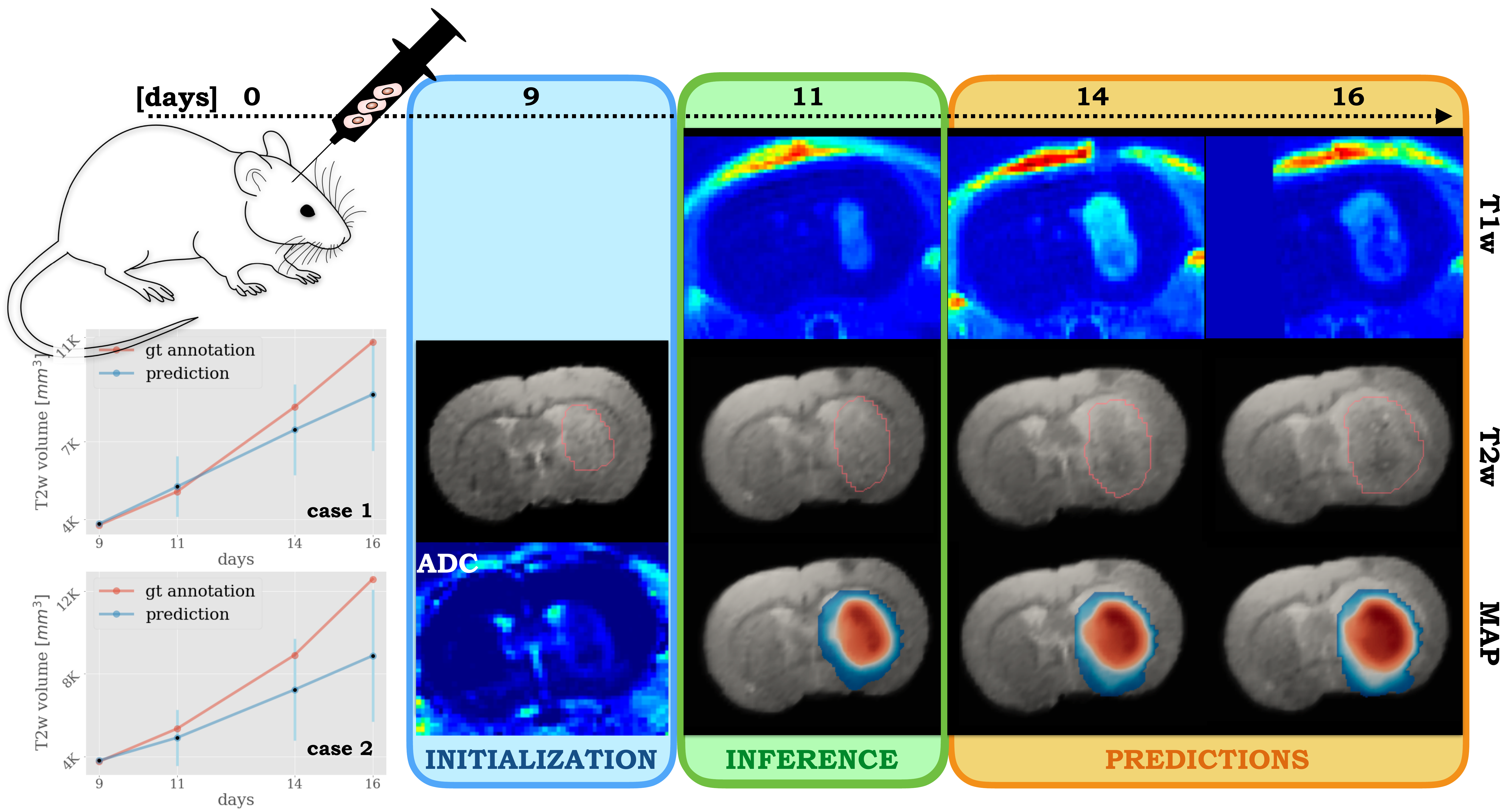}
\caption{\small{Data preparation and results of the inference for the real rat cases. After the injection of cancer lines, the rats are monitored at four time points by means of the T1w, T2w and DWI modalities. The inverse ADC from the DWI at day 9 serves as an initial condition for the tumor model. The model parameters are inferred from the binary segmentations, obtained from the T1w and T2w scans at day 11. The model predictions, simulated using MAP estimations of the $D,\rho$ parameters, are validated at days 14 and 16. In the bottom row, the predicted tumor cell density profiles, overlayed on the T2w, are shown. In the middle row, the pink outlines are boundaries of the T2w segmentations. The inset shows volume dynamics of the T2w binary segmentations, obtained from the annotation and predicted by the model, for two rats (the rat case 1 corresponds to the scans shown above). The inferred parametric uncertainties were propagated through the model to obtain mean and standard deviation (blue bars) of the dynamics.}} \label{fig3}
\end{figure}

We validated the method on two real rat cases. Fig. \ref{fig3} shows the tumor cell distributions for one of the rats, 
\noindent simulated using the inferred MAP estimates, at the calibration time point (day 11) and at the validation points (days 14 and 16).The volume dynamics of T2 tumor segmentations for both rats, calculated from the annotation and predicted by the model, is shown on the inset. The predicted segmentations were obtained by thresholding the simulated cell density profile at the inferred $u^{T2}$ level. While the volumes are in good agreement at the calibration time (an indication of plausible parametric inference), the model prediction underestimates the real tumor dynamics at the validation points. This can be attributed to simplifying assumptions of the Fisher-Kolmogorov model, such as constant proliferation and diffusion for all time points, which limit model's ability to describe the nonlinear character of the real tumor progression. As the proposed inference scheme opens an avenue for efficient parametric estimation, we will test more complicated tumor growth formalisms, e.g., accounting for tissue displacement and microenviromental influence, in future work.
\section{Conclusion}
We present an approach for inferring parameters of a tumor model from information available on medical scans relying on a learning-based strategy. The approach allows for more efficient parametric estimation, as compared to the conventional Bayes description with MCMC type of sampling, and exempts from the necessity to introduce an explicit form, linking the biophysical model with image observation. Despite we demonstrate the applicability of the method to tumor modeling, the method can be adopted to other physical modeling problems that require calibration from imaging modalities.\\

\noindent\textbf{Acknowledgment:} Ivan Ezhov and Suprosanna Shit have received funding from the European Union’s Horizon 2020 research and innovation programme under the Marie Sklodowska-Curie grant agreement TRABIT No 765148.

\bibliographystyle{splncs.bst}
\bibliography{mybibliography}

\end{document}